\documentclass[sn-mathphys]{sn-jnl}

\jyear{2021}%

\usepackage{enumerate}
\usepackage{float}
\usepackage{graphicx}
\usepackage{subfigure}
\usepackage{algorithm}

\newcommand{\bra}[1]{ \left\lvert  #1\right\rangle}

\theoremstyle{thmstyleone}%
\theoremstyle{thmstyletwo}%

\theoremstyle{thmstylethree}%
\newtheorem{definition}{Definition}%

\begin{document}

\title[Li et al.]{Quantum Privacy-preserving Two-party Circle Intersection Protocol Based on Phase-encoded Query}


\author[1]{\fnm{Zi-Xian} \sur{Li}}\email{zixianli157@163.com}
\equalcont{These authors contributed equally to this work.}

\author[1]{\fnm{Qi} \sur{Yang}}\email{yangqi7275@163.com}
\equalcont{These authors contributed equally to this work.}

\author[2]{\fnm{Bao} \sur{Feng}}\email{fengbao@sgepri.sgcc.com.cn}
\equalcont{These authors contributed equally to this work.}

\author*[1,3]{\fnm{Wen-Jie} \sur{Liu}}\email{wenjiel@163.com}

\affil*[1]{\orgdiv{School of Software}, \orgname{Nanjing University of Information Science and Technology}, \orgaddress{\street{No.~219 Ningliu Road}, \city{Nanjing}, \postcode{210044}, \state{Jiangsu}, \country{China}}}

\affil[2]{\orgname{NARI Information \& Communication Technology Co., Ltd.}, \orgaddress{\street{No.~8 Nanrui Road}, \city{Nanjing}, \postcode{210003}, \state{Jiangsu}, \country{China}}}

\affil[3]{\orgdiv{Engineering Research Center of Digital Forensics}, \orgname{Ministry of Education}, \orgaddress{\street{No.~219 Ningliu Road}, \city{Nanjing}, \postcode{210044}, \state{Jiangsu}, \country{China}}}


\abstract{Privacy-preserving geometric intersection (PGI) is an important issue in Secure multiparty computation (SMC). The existing quantum PGI protocols are mainly based on grid coding, which requires a lot of computational complexity. The phase-encoded query method which has been used in some Quantum SMC protocols is suitable to solve the decision problem, but it needs to apply high dimensional Oracle operators. In this paper, we use the principle of phase-encoded query to solve an important PGI problem, namely privacy-preserving two-party circle intersection. We study the implementation of Oracle operator in detail, and achieve polynomial computational complexity by decompsing it into quantum arithmetic operations. Performance analysis shows that our protocol is correct and efficient, and can protect the privacy of all participants against internal and external attacks.}




\maketitle

\section{Introduction}\label{sec1}

Since Yao \cite{Yao1982} proposed the concept of secure multiparty computation (SMC) in 1982, it has quickly become an important and fruitful research field in modern cryptography. In SMC, two or more participants, using their own secret information as input, work together to perform confidential calculations while ensuring that each party only obtains the calculated results it should obtain. Because classical cryptography would be in face of attacks from quantum computer, some scholars \cite{Shi2015,Ji2019,Liu2019B,Shi2021,Liu2022,Ye2022} use quantum mechanism to achieve SMC, namely quantum SMC (QSMC), to achieve higher security or computational efficiency than classical SMC.

Privacy-preserving computational geometry (PCG) is a kind of important SMC problem, which mainly considers the calculation of position relations between several nodes scattered in different spatial locations. For example, calculate the distance \cite{Atallah2001,Huang2016}, azimuth \cite{Chen2018} or included angle \cite{Chen2018} between some geometric objects. In the field of QSMC, researches on PCG mainly focus on geometric intersection decision, such as privacy-preserving geometric intersection (PGI) \cite{Liu2019}, point inclusion \cite{Shi2017}, etc, or two-party distance \cite{Peng2017,Peng2018,Chen2018,Cao2022} and scalar product \cite{He2012,Shi2019} computation, which are relatively simple computation problems. Among them, PGI mainly studies how to determine intersecting relations between several geometric areas, which is a hotspot in the classical field \cite{Atallah2001,Li2014,Zhu2018}. In Quantum PGI (QPGI), one solution is to represent several private points and areas as a set of numbered grids, which are not constrained by specific area shapes and can be easily implemented using existing technologies. In 2016, Shi et al. \cite{Shi2017} first proposed a quantum point inclusion protocol based on phase-encoded query \cite{Olejnik2011}. The protocol successfully uses quantum advantage to determine whether one party Alice's private point belongs to the other party Bob's private area with much lower communication complexity than classical computation. However, it is difficult to implement the phase-encoded query because it needs to implement an Oracle operator in high dimensions. In 2019, Liu et al. \cite{Liu2019} used the quantum counting algorithm (a variant of the famous Grover's algorithm \cite{Grover1997}) to realize the privacy-preserving intersection decision of two private geometric areas by grid-coding. However, since Grover's algorithm can only realize square acceleration, its computational efficiency is not high enough.

In general, the computational complexity of grid-coding-based methods often depends linearly on the number of grid points, rather than the size of private areas. It means that even in the case of two very small areas on the plane, grid-coding methods often need to determine all the grid points, which greatly affects the computational efficiency. Privacy-preserving circle intersection (PCI) is a kind of representative PGI problem. Because circle areas have regular geometry structure, their intersection can be determined by calculating the center coordinates and radius of the circles, without coding every grid. In this paper, we use the principle of phase-encoded query to realize the quantum privacy-preserving two-party circle intersection decision. We study the implementation of Oracle operator in detail, and decompose it into several quantum arithmetic operations. By this way, the computational complexity of polynomial level is achieved, which avoids the problem that high dimensional Oracle operator is difficult to achieve. We also analyze the performance of our protocol, and prove that it is correct and efficient, and can protect the privacy of all participants against internal and external attacks.

The rest of this paper is arranged as follows: In Section~\ref{sec2}, we do some preliminary work, including showing some basic quantum gates, reviewing the phase-encoded query protocol and introducing some required quantum arithmetic operations. In Section~\ref{sec3}, we present our protocol in detail, including the implementation of the Oracle operator and the specific process of the protocol. We analyze the proposed protocol in Section~\ref{sec4} and conclude in Section~\ref{sec5}.

\section{Preliminary}\label{sec2}

\subsection{Basic quantum gate}\label{sec2.1}
The basic quantum gates we will use are shown here.

\begin{enumerate}[(1)]
    \item {Pauli X gate $\sigma_X=\begin{bmatrix} 0 & 1\\1&0\end{bmatrix}$:
    \[\sigma_X: \bra{a}\to \bra{a\oplus 1}, a=0,1,\] 
    where ``$\oplus$'' means XOR.}
    \item {Pauli Z gate $\sigma_Z=\begin{bmatrix} 1 & 0\\0&-1\end{bmatrix}$:
    \[\sigma_X: \bra{a}\to (-1)^{a}\bra{a}, a=0,1.\]}
    \item {Hadamard gate $H=\frac{1}{\sqrt{2}}\begin{bmatrix} 1 & 1\\1&-1\end{bmatrix}$:
    \[H: \bra{0}\to \bra{+}=\frac{\bra{0}+\bra{1}}{\sqrt{2}},\bra{1}\to \bra{-}=\frac{\bra{0}-\bra{1}}{\sqrt{2}}.\]}
    \item {Controlled $\sigma_X$ (CNOT) gate $CNOT=\begin{bmatrix} 1 & 0&0&0\\0&1&0&0\\0&0&0&1\\ 0&0&1&0\end{bmatrix}$:
    \[CNOT: \bra{a}\bra{b}\to \bra{a}\bra{b\oplus a}, a,b=0,1.\]}
\end{enumerate}

\subsection{Phase-encoded query protocol review}\label{sec2.2}
Phase-encoded query protocol \cite{Olejnik2011}: Assume that Bob has a database of single-bit numbers, where all the data are numbered as $1,...,N - 1$. Alice wants to query Bob's $i$-th data ${x_i}$, but does not want Bob to know the query index $i$.

To achieve the above query, let $m = \left \lceil {\log _2}N \right \rceil $ be the number of qubits. Alice encodes her input $i$ as a $m$-qubit superposition state as
\[\frac{\bra{0}+ \bra{i}}{\sqrt 2}.\]
 Although this state involves entanglement of $m$ qubits, it can be generated by using $m$ times controlled $\sigma_X$ gates ($CNOT$). Note that 
\[\frac{\bra{0}+ \bra{i}}{\sqrt 2}= \frac{\bra{0}\bra{0}\cdots\bra{0}+ \bra{i_{m-1}}\bra{i_{m-2}}\cdots\bra{i_{0}}}{\sqrt 2}.\]
Assume that the $m$-bit number $i = {i_{m - 1}}{i_{m - 2}} \ldots {i_1}{i_0}$ has $k$ non-zero bits numbered ${j_1},{j_2},...,{j_k}$. Then Alice first generates the state 
\[\bra{+}_{{j_1}}=\frac{\bra{0}_{j_1} +\bra{1}_{j_1}}{\sqrt 2 } \]
by applying Hadamard gate $H$ on $\bra{0}_{j_1}$, and then performs a unitary transformation 
\[\prod\limits_{x = 1}^k {CNOT^{\left( {{j_1},{j_x}} \right)}}: \bra{+}_{j_1}\bra{0}^{\otimes (m-1)} \to\frac{\bra{0}\bra{0}\cdots\bra{0}+ \bra{i_{m-1}}\bra{i_{m-2}}\cdots\bra{i_{0}}}{\sqrt 2},\]
where $CNOT^{\left( {{j_1},{j_x}} \right)}$ copies the first non-zero bit $i_{j_1}$ to the other non-zero $i_{j_x}$. Thus, the preparation of the high-dimensional entangled state $\frac{\bra{0}+ \bra{i}}{\sqrt 2}$ is realized. 

\begin{figure}
\centering
\includegraphics[width=0.8\textwidth]{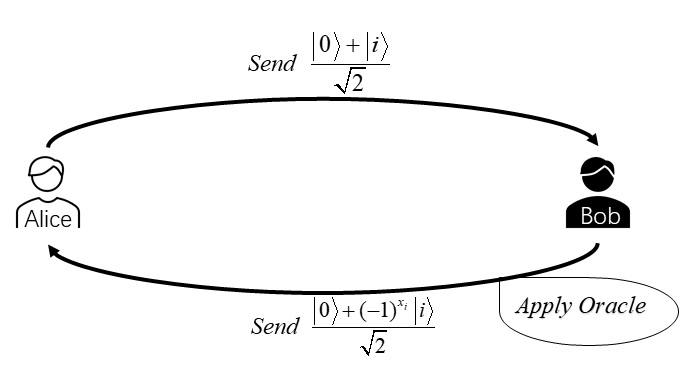}
\caption{A diagrammatic sketch of phase-encoded query.}\label{fig1}
\end{figure}

As shown in Fig~\ref{fig1}, Alice now sends her quantum state to Bob via a verified quantum channel. After Bob receives the status, he applies an Oracle operator: 
\[U = \left( {\begin{array}{*{20}{c}}
1&0& \ldots &0\\
0&{{{\left( { - 1} \right)}^{{x_1}}}}& \ldots &0\\
 \vdots & \vdots & \ddots & \vdots \\
0&0& \ldots &{{{\left( { - 1} \right)}^{{x_{N - 1}}}}}
\end{array}} \right).\]
It performs as
\[U:\frac{\bra{0}+ \bra{i}}{\sqrt 2}\to \frac{\bra{0}+(-1)^{x_i}\bra{i}}{\sqrt 2},\]
by a conditional phase flip. Bob then sends this state to Alice. 

To extract the solution ${x_i}$ from the phase-encoded state,
Alice must distinguish between two possible states 
\[\bra{i_+}=\frac{\bra{0}+\bra{i}}{\sqrt 2},\ \ 
\bra{i_-}=\frac{\bra{0}-\bra{i}}{\sqrt 2}.
\]
In general, the distinguishing is described as a projection measurement performed on the basis $\bra{i_+},\bra{i_-}$, with projection operators ${P_+} = \left\lvert {i_+} \right\rangle \left\langle {{i_+}} \right\lvert$ and ${P_-} = \left\lvert {{i_-}} \right\rangle \left\langle {{i_-}} \right\lvert$. Although the measurement involves bases in high dimensions Hilbert space, it can be converted into a measurement on single qubit. This can also be done using $m$ times $CNOT$ gates. By a process exactly opposite to the preparation of $\bra{\psi_1}$, Alice can change the state $\bra{\psi_2}$ to form $\bra{\pm}\bra{0}^{\otimes (m-1)}$ and measure the single qubit under the basis $\bra{+}$ and $\bra{-}$ to distinguish the received state. Depending on whether the measurement is $\bra{+}$ or $\bra{-}$, Alice will know whether ${x_i} = 0$ or $1$.

Bob cannot know Alice's query index $i$ with certainty throughout the computation. He could choose to measure the state $\frac{\bra{0}+\bra{i}}{\sqrt 2}$ halfway through and get $\bra{i}$ with probability $\frac{1}{2}$, but he would not succeed $100\%$. However, the phase-encoded query protocol still has a certain probability of information leakage. Shi et al. \cite{Shi2017} design a honesty test that allows Alice to detect Bob's cheating with a certain probability, which will be used in our protocol.

\subsection{Quantum arithmetic operation}\label{sec2.3}
In addition to the basic quantum gates described in Section~\ref{sec2.1}, several quantum arithmetic operations need to be introduced. At first, for each two $n$-bit integers $x,y$, the following semi-quantum modular adder is needed:
\[{U_{+x\bmod {2^n}}}:\bra{y} \to \bra{y + x\bmod {2^n}}.\]
It has a full quantum version as:
\[U_{ + \bmod {2^n}}^{(12)}:{\bra{x}_1}{\bra{y}_2} \to \bra{x}_1\bra{y + x\bmod {2^n} }_2.\]
As a quantum extension of classical modular adders, an $n$-qubit quantum modular adder requires $O(n)$ qubits and $O\left( n \right)$ computational complexity.

We also need a semi-quantum modular multiplier as
\[U_{ \times x\bmod {2^n}}^{(12)}:\bra{y}_1 \bra{c}_2\to \bra{y}_1\bra{c + xy\bmod {2^n}}_2,\]
which was used to realize Shor's factoring algorithm \cite{Shor1997}. Similar, it has a full quantum version as
\[U_{ \times \bmod {2^n}}^{\left( {123} \right)}:\left\lvert {{{\left.x \right\rangle }_1}} \right.\left\lvert {{{\left.y \right\rangle }_2}} \right.\left\lvert {{{\left.c \right\rangle }_3}} \right.\to \left\lvert {{{\left.x \right\rangle }_1}} \right.\left\lvert {{{\left.y \right\rangle }_2}} \right.\left\lvert {{{\left.{c + xy\bmod {2^n}} \right\rangle }_3}} \right.\]
A quantum modular multiplier can be done by taking each qubit $y_i$ ($i = 0,1,...,n - 1$) of quantum integer ${\bra{y}}$ as the control qubit to add up ${2^i}x\bmod {2^n}$ (i.e., taking the highest $i$ qubits of $x$) to $\bra{c}_3$. Since each step is one run of quantum modular adder, the computational complexity is $O\left(n^2 \right)$.

\section{Quantum privacy-preserving two-party circle intersection protocol}\label{sec3}

\begin{figure}
\centering
\includegraphics[width=0.8\textwidth]{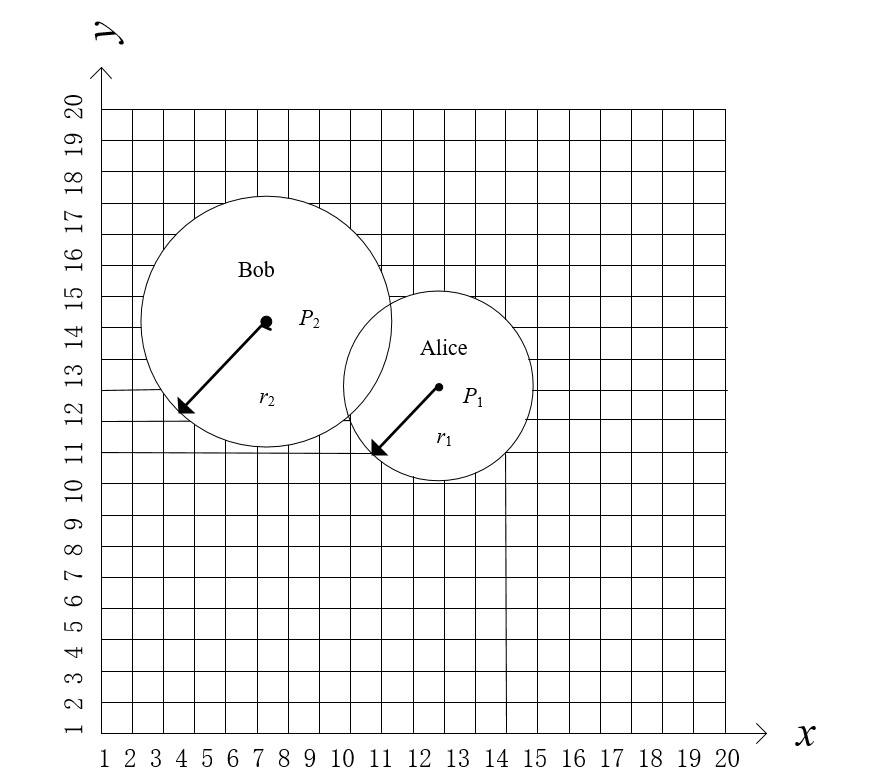}
\caption{A diagrammatic sketch of P2CI problem.}\label{fig2}
\end{figure}

\begin{definition}[Privacy-preserving two-party circle intersection, P2CI] As shown in Fig~\ref{fig2}, given a 2-dimensional background plane with a coordinate system $x,y \in \left\{ {1,...,T - 1} \right\}$, where the size of $T = {2^t}$ can be determined according to precision requirements. Alice and Bob each have a private plane circle area $A$ and $B$ with centers ${P_1} = \left( {{x_1},{y_1}} \right)$ and ${P_2} = \left( {{x_2},{y_2}} \right)$ and radii ${r_1}$ and ${r_2}$, respectively. The two parties want to determine if areas $A$ and $B$ intersect, without revealing any other information about their own area to each other.
\end{definition}

The basic idea is to use the necessary and sufficient conditions for circles to intersect, i.e.,
\[d = \sqrt {{{\left( {{x_1} - {x_2}} \right)}^2} + {{\left( {{y_1} - {y_2}} \right)}^2}}  < {r_1} + {r_2}.\]
Since the input must be an integer, we square it:
\[D = {\left( {{x_1} - {x_2}} \right)^2} + {\left( {{y_1} - {y_2}} \right)^2} < {\left( {{r_1} + {r_2}} \right)^2} = R.\]
Then, the phase-encoded query can be used in such determination problem, because it can easily send single bit data. Firstly, we need to use quantum arithmetic operations to realize the Oracle operator of phase-encoded query, and then use phase-encoded query to realize circle intersection decision.

\subsection{Oracle operator implementation}\label{sec3.1}
To implement the Oracle operator, first we use $n$ bits to represent any integer $ - {2^{n - 1}} \le y \le {2^{n - 1}} - 1$, i.e., 
\[y \to \bra{y\bmod {2^n}} = \left\{ {\begin{array}{*{20}{c}}
{\bra{y} ,0 \le y \le {2^{n - 1}} - 1}\\
{\bra{2^n+y} , - {2^{n - 1}} \le y < 0}
\end{array}} \right..\]
Obviously the mapping is one-to-one, where case $y < 0$ means $y\bmod {2^n} > {2^{n - 1}}$, so the highest bit of $y$ is 1, otherwise 0. We now need a conditional phase flipping operator to determine the size relationship between quantum integer $\bra{y}$ and classical integer $0$, i.e.,
\[{O^{\left( {y < 0} \right)}}:\bra{y} \to (- 1)^{g(y )}\bra{y},\]
where
\[g\left( y \right) = \left\{ {\begin{array}{*{20}{c}}
{1,y < 0}\\
{0,y \ge 0}
\end{array}} \right..\]
Obviously this phase flipping can be achieved by applying the Pauli $Z$ gate $\sigma_Z:\bra{a} \to (-1)^a\bra{a}$ to the highest qubit $\bra{y_{n - 1}}$. When $y < 0$, $\bra{y_{n - 1}}$ is $\bra{1}$, so the phase will flip. Otherwise, it won't flip. So we have \[{O^{\left( {y < k} \right)}} = {\sigma_Z^{(n - 1)}}.\]

Let Alice's query input be $X = {x_1}\parallel {y_1}\parallel {r_1}$, where each data ${x_1},{y_1},{r_1}$ is a $t$-bit integer. For 
\[\begin{aligned}
&D = {\left( {{x_1} - {x_2}} \right)^2} +{\left( {{y_1} - {y_2}} \right)^2}\\
&R = {\left( {{r_1} + {r_2}} \right)^2},\end{aligned}\]
we need to implement the following Oracle operator 
\[U:\bra{X}\to (-1)^{f(X)}\bra{X},\]
where $f\left( {{x_1},{y_1},{r_1}} \right) = \left\{ {\begin{array}{*{20}{c}}
{1,D - R < 0}\\
{0,D - R \ge 0}
\end{array}} \right..$
Note that 
\[\begin{aligned}
    &D = {\left( {{x_1} - {x_2}} \right)^2} + {\left( {{y_1} - {y_2}} \right)^2} < {2^{2t}} + {2^{2t}} = {2^{2t + 1}}\\
    &R = {\left( {{r_1} + {r_2}} \right)^2} < {\left( {{2^{t + 1}}} \right)^2} = {2^{2t + 2}}\\
    &\left\lvert {D - R} \right\lvert \le \max \left( {D,R} \right) < {2^{2t + 2}},
\end{aligned}\] 
and 
\[\begin{aligned}
&\ \ \ D - R \\
&\equiv {\left( {{x_1} - {x_2}} \right)^2} + {\left( {{y_1} - {y_2}} \right)^2} - {\left( {{r_1} + {r_2}} \right)^2}\\
 &\equiv {x_1}^2 + {y_1}^2 - {r_1}^2 + {x_1} \cdot \left( { - 2{x_2}} \right) + {y_1} \cdot \left( { - 2{y_2}} \right) + {r_1}\left( { - {2r_2}} \right) + \left( {{x_2}^2 + {y_2}^2 - {r_2}^2} \right)\\
 &\equiv {x_1}^2 + {y_1}^2 - {r_1}^2 + {x_1} \cdot \left( {{2^n} - 2{x_2}} \right) + {y_1} \cdot \left( {{2^n} - 2{y_2}} \right) + {r_1} \cdot \left( {{2^n} - {2r_2}} \right) +\\
 &\left( {{x_2}^2 + {y_2}^2 - {r_2}^2} \right)(\bmod {2^n}).
\end{aligned}\]
If taking \[\begin{aligned}
&{k_1} = \left( {{2^n} - 2{x_2}} \right)\bmod {2^n}\\
&{k_2} = \left( {{2^n} - 2{y_2}} \right)\bmod {2^n}\\
&{k_3} = \left( {{2^n} - {2r_2}} \right)\bmod {2^n}\\
&{k_4} = {2^n} - \left( {{x_2}^2 + {y_2}^2} \right)\bmod {2^n},\end{aligned}\]
then \[D \equiv {x_1}^2 + {y_1}^2 - {r_1}^2 + {x_1} \cdot {k_1} + {y_1} \cdot {k_2} + {r_1} \cdot {k_3} + {k_4}(\bmod {2^n}).\]
So we set $n = 2t + 3$, and use $n$-qubit registers to store all the integers. By using $n$-qubit modular adder and multiplier, we can determine whether $D-R$ is greater than 0. 

The specific process is as follows. Let the input be $3n$-qubit register $\bra{X}  = \bra{x_1}_1\bra{y_1}_2\bra{r_1}_3$. Firstly, a single qubit auxiliary register ${\bra{1}_a}$ is taken, and then all qubits of $\bra{X}$ are reversed by applying $\sigma_X$ gates. Then take each qubit of it as the control qubits to impose a controlled $\sigma_X$ gate on ${\bra{1}_a}$. Thus,
\[\bra{1}_a \to \left\{ {\begin{array}{*{20}{c}}
{\bra{0}_a,X = 0}\\
{\bra{1}_a,X \ne 0}
\end{array}} \right..\]
Now reverse all qubits of $\bra{X}$ again to recover it. At this point, the following operations are all performed controlled by register $a$ (which means that if $\bra{X}  = \bra{0}$, no operation will be performed, and if $\bra{X}  \ne \bra{0} $ then perform the following operations):
\begin{enumerate}[Step 1]
    \item {Prepare three $n$-qubit auxiliary registers ${e_1},{e_2},{e_3}$, then apply $CNOT^{ \otimes n}$ gates:
\[\begin{aligned}
    &CNOT^{ \otimes n}:{\left\lvert {{x_1}} \right\rangle _1}{\left\lvert 0 \right\rangle _{{e_1}}} \to {\left\lvert {{x_1}} \right\rangle _1}{\left\lvert {{x_1}} \right\rangle _{{e_1}}}\\
    &CNOT^{ \otimes n}:{\left\lvert {{y_1}} \right\rangle _2}{\left\lvert 0 \right\rangle _{{e_2}}} \to {\left\lvert {{y_1}} \right\rangle _2}{\left\lvert {{y_1}} \right\rangle _{{e_2}}}\\
    &CNOT^{ \otimes n}:{\left\lvert {{r_1}} \right\rangle _3}{\left\lvert 0 \right\rangle _{{e_3}}} \to {\left\lvert {{r_1}} \right\rangle _3}{\left\lvert {{r_1}} \right\rangle _{{e_3}}}.\end{aligned}\]}
    \item {Take the complement of $\bra{r_1}_3$ (i.e., qubit-wise reverse it first, and then modular add 1 to it):
\[U_{ + 1\bmod {2^n}}{{\sigma_X}^{\otimes n}}:\bra{r_1}_3\to \bra{2^n-r_1}_3=\bra{-r_1}_3.\]}

    \item {Prepare another three $n$-qubit auxiliary registers ${g_1},{g_2},{g_3}$ for full quantum modular multiplier: 
\[\begin{aligned}
&\ \ \ \ \ \ {U_{ \times \bmod {2^n}}}:\\
&{\left\lvert {{x_1}} \right\rangle _1}{\left\lvert {{x_1}} \right\rangle _{{e_1}}}{\left\lvert 0 \right\rangle _{{g_1}}} \to {\left\lvert {{x_1}} \right\rangle _1}{\left\lvert {{x_1}} \right\rangle _{{e_1}}}{\left\lvert {{x_1}^2} \right\rangle _{{g_1}}}\\
&{\left\lvert {{y_1}} \right\rangle _2}{\left\lvert {{y_1}} \right\rangle _{{e_2}}}{\left\lvert 0 \right\rangle _{{g_2}}} \to {\left\lvert {{y_1}} \right\rangle _2}{\left\lvert {{y_1}} \right\rangle _{{e_2}}}{\left\lvert {{y_1}^2} \right\rangle _{{g_2}}}\\
&{\left\lvert { - {r_1}} \right\rangle _3}{\left\lvert {{r_1}} \right\rangle _{{e_3}}}{\left\lvert 0 \right\rangle _{{g_3}}} \to {\left\lvert { - {r_1}} \right\rangle _3}{\left\lvert {{r_1}} \right\rangle _{{e_3}}}{\left\lvert { - {r_1}^2} \right\rangle _{{g_3}}},\end{aligned}\]}

    \item {Apply $U_{ + 1\bmod {2^n}}{{\sigma_X}^{\otimes n}}$ again to restore $\bra{-r_1}_3$ to $\bra{r_1}_3$. Apply $CNO{T^{ \otimes n}}$ gate again to restore ${e_1},{e_2},{e_3}$ to $\bra{0}$. Then perform semi-quantum modular multiplier:
\[\begin{aligned}
&{U_{ \times {k_1}\bmod {2^n}}}:{\left\lvert {{x_1}} \right\rangle _1}{\left\lvert 0 \right\rangle _{{e_1}}} \to {\left\lvert {{x_1}} \right\rangle _1}{\left\lvert {{x_1}{k_1}} \right\rangle _{{e_1}}}\\
&{U_{ \times {k_2}\bmod {2^n}}}:{\left\lvert {{y_1}} \right\rangle _2}{\left\lvert 0 \right\rangle _{{e_2}}} \to {\left\lvert {{y_1}} \right\rangle _2}{\left\lvert {{y_1}{k_2}} \right\rangle _{{e_2}}}\\
&{U_{ \times {k_3}\bmod {2^n}}}:{\left\lvert {{r_1}} \right\rangle _3}{\left\lvert 0 \right\rangle _{{e_3}}} \to {\left\lvert {{r_1}} \right\rangle _3}{\left\lvert {{r_1}{k_3}} \right\rangle _{{e_3}}},\end{aligned}\]}

    \item {Apply full quantum modular adder:
\[\begin{aligned}
&\ \ \ \ {U_{ + \bmod {2^n}}}:\\
&{\left\lvert {{x_1}{k_1}} \right\rangle _{{e_1}}}{\left\lvert {{x_1}^2} \right\rangle _{{g_1}}} \to {\left\lvert {{x_1}{k_1}} \right\rangle _{{e_1}}}{\left\lvert {{x_1}^2 + {x_1}{k_1}} \right\rangle _{{g_1}}},\\
&{\left\lvert {{y_1}{k_2}} \right\rangle _{{e_2}}}{\left\lvert {{y_1}^2} \right\rangle _{{g_2}}} \to {\left\lvert {{y_1}{k_2}} \right\rangle _{{e_2}}}{\left\lvert {{y_1}^2 + {y_1}{k_2}} \right\rangle _{{g_2}}},\\
&{\left\lvert {{r_1}{k_3}} \right\rangle _{{e_3}}}{\left\lvert { - {r_1}^2} \right\rangle _{{g_3}}} \to {\left\lvert {{r_1}{k_3}} \right\rangle _{{e_3}}}{\left\lvert { - {r_1}^2 + {r_1}{k_3}} \right\rangle _{{g_3}}},\\
&{\left\lvert {{x_1}^2 + {x_1}{k_1}} \right\rangle _{{g_1}}}{\left\lvert { - {r_1}^2 + {r_1}{k_3}} \right\rangle _{{g_3}}} \to {\left\lvert {{x_1}^2 + {x_1}{k_1}} \right\rangle _{{g_1}}}{\left\lvert {{x_1}^2 - {r_1}^2 + {x_1}{k_1} + {r_1}{k_3}} \right\rangle _{{g_3}}},\\
&{\left\lvert {{y_1}^2 + {y_1}{k_2}} \right\rangle _{{g_2}}}{\left\lvert {{x_1}^2 - {r_1}^2 + {x_1}{k_1} + {r_1}{k_3}} \right\rangle _{{g_3}}}\\
&\ \ \ \ \ \to {\left\lvert {{y_1}^2 + {y_1}{k_2}} \right\rangle _{{g_2}}}{\left\lvert {{x_1}^2 + {y_1}^2 - {r_1}^2 + {x_1}{k_1} + {y_1}{k_2} + {r_1}{k_3}} \right\rangle _{{g_3}}}.\end{aligned}\]}

    \item {Apply semi-quantum modular adder:
\[\begin{aligned}
&{U_{+{k_4}\bmod {2^n}}}:{\left\lvert {{x_1}^2 + {y_1}^2 - {r_1}^2 + {x_1}{k_1} + {y_1}{k_2} + {r_1}{k_3}} \right\rangle _{{g_3}}}\\
& \ \ \ \ \ \to {\left\lvert {{x_1}^2 + {y_1}^2 - {r_1}^2 + {x_1}{k_1} + {y_1}{k_2} + {r_1}{k_3} + {k_4}} \right\rangle _{{g_3}}} \\
&\ \ \ \ \ \ = {\left\lvert {D - R} \right\rangle _{{g_3}}}.
\end{aligned}\]}

    \item {Apply the controlled phase flipping operator:
\[{O^{\left( {D - R < 0?} \right)}}:{\left\lvert {D - R} \right\rangle _{{g_3}}} \to {\left( { - 1} \right)^{g\left( {D - R} \right)}}{\left\lvert {D - R} \right\rangle _{{g_3}}},\] 
where 
\[g\left( {D - R} \right) = \left\{ {\begin{array}{*{20}{c}}
{1,D - R < 0}\\
{0,D - R \ge 0}
\end{array}} \right..\]}

    \item {Inverse everything we did in Step 1-6 (denoted as $U_{D-R}$) to recover all the registers to their original states.}

\end{enumerate}

\begin{figure} \centering
\subfigure[The overall circuit, where $\bra{0}_4=\bra{0}_{e_1}\bra{0}_{e_2}\bra{0}_{e_3}\bra{0}_{g_1}\bra{0}_{g_2}\bra{0}_{g_3}.$] { \label{fig3a}
\includegraphics[width=0.8\columnwidth]{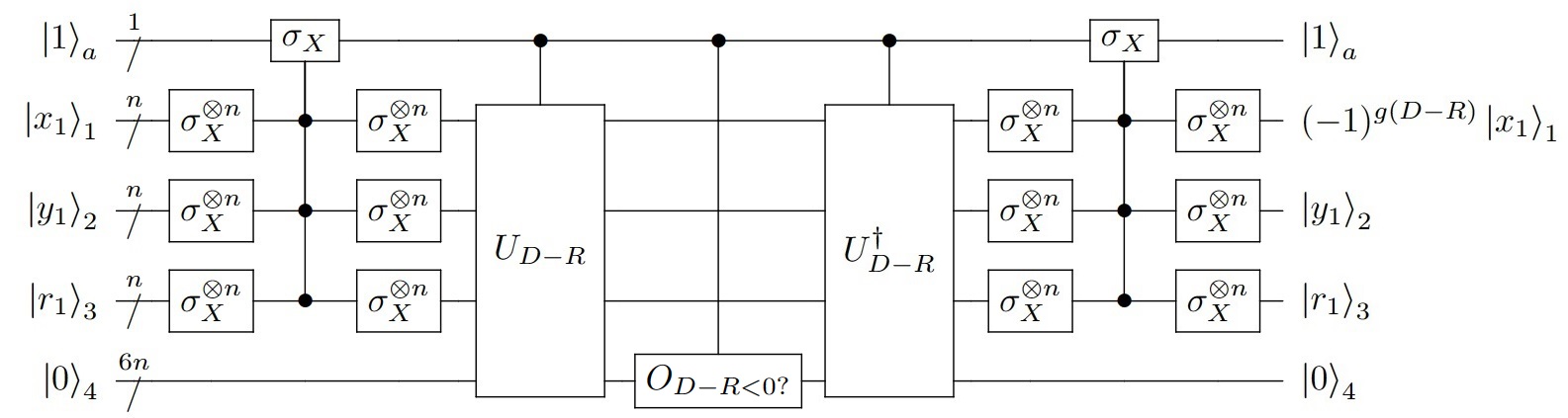}
}
\subfigure[The detail circuit of $U_{D-R}$.] { \label{fig3b}
\includegraphics[width=1\columnwidth]{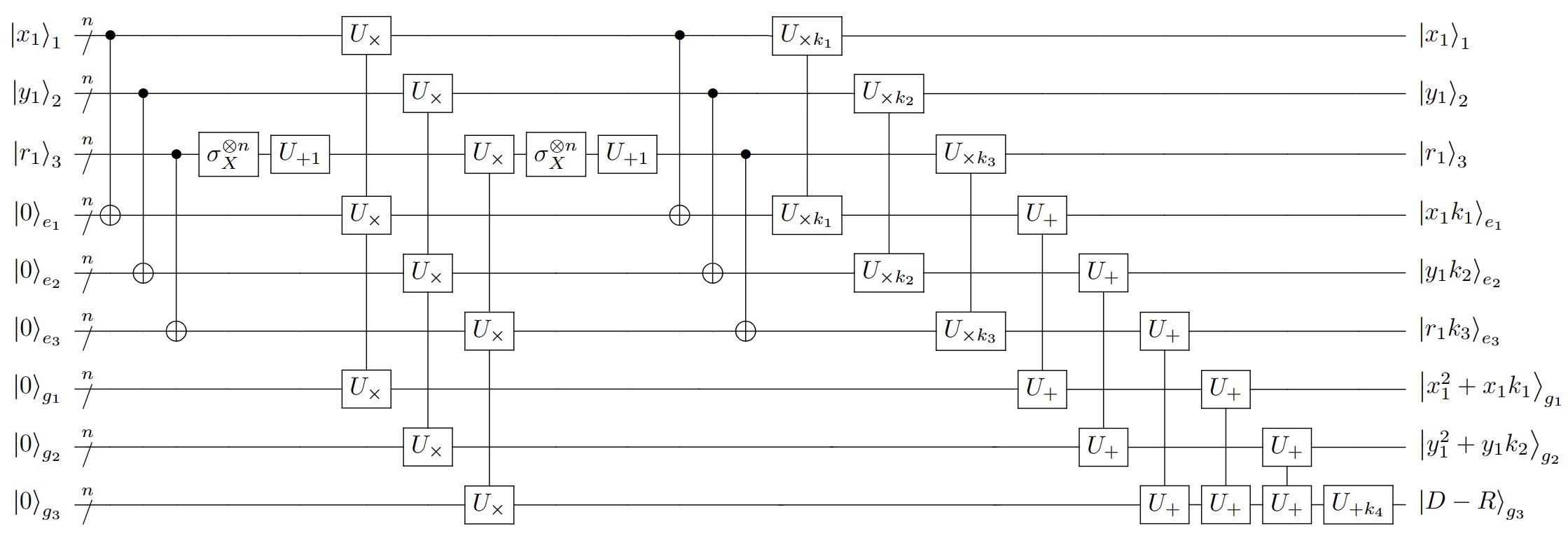}
}
\caption{The circuit implementation of the Oracle operator.}
\label{fig3}
\end{figure}

Finally, we need apply controlled $\sigma_X$ gate to recover register $a$ to $\bra{1}$. Now, except the phase $(-1)^{g(D-R)}$, everything is the same as the beginning. In this way, we implement the unitary operator $U$ that we need. The circuit diagram of the entire process is shown in Fig~\ref{fig3a} and~\ref{fig3b}.

\subsection{Protocol process}\label{sec3.2}
We provide the specific process of our protocol as follows.

\subsubsection{Preparation stage}\label{sec3.2.1}
\begin{enumerate}[Step 1]
    \item {Alice prepares two $m = 3n$-qubit particles ${h_1}$, ${t_1}$ initialed as state $\bra{0}$. She now 
 applies $\prod\limits_{x = 1}^k {CNOT^{\left( {{j_1},{j_x}} \right)}}H^{j_1}$ (as described in Section~\ref{sec2.2}) on $h_1$, to change $\bra{0}_{h_1}$ to $\frac{\bra{0}_{h_1}+\bra{X}_{h_1}}{\sqrt 2 }$, where $X = {x_1}\parallel {y_1}\parallel {r_1}$ is her information string of the circle area.}
 
 \item {Alice then applies $m$ times $CNOT$ gates to entangle the two particles ${h_1},{t_1}$, where the $i$-th qubit of ${h_1}$ is taken as the control qubit and the corresponding $i$-th qubit of ${t_1}$ is taken as the target qubit. Now the state is 
\[ CNOT^{\otimes m}:\frac{\bra{0}_{h_1}+ \bra{X}_{h_1}}{\sqrt 2 } \bra{0}_{t_1}\to \frac{\bra{0}_{h_1}\bra{0}_{t_1}+ \bra{X}_{h_1}\bra{X} _{t_1}}{\sqrt 2 }.\]}

    \item {Further, Alice prepares another two $m$-qubit particles ${h_2},{t_2}$ in state 
\[\frac{\bra{0}_{h_2}\bra{0}_{t_2}+\bra{X} _{h_2}\bra{X}_{t_2}}{\sqrt 2 }.\]
The preparation is the same as Step 1.}

    \item {Bob construct the Oracle operator $U$ as described in Section~\ref{sec3.1}.}
\end{enumerate}

\subsubsection{Operation stage}\label{sec3.2.2}
\begin{enumerate}[Step 1]
    \item Alice sends particles ${t_1}$ and ${t_2}$ to Bob, while keeping particles ${h_1}$ and ${h_2}$ in her hand.
    \item {After receiving particles ${t_1}$ and ${t_2}$, Bob applies the Oracle operator $U$ on ${t_1}$ and ${t_2}$ respectively: 
\[U:\frac{\bra{0}_{h_i}\bra{0}_{t_i}+\bra{X}_{h_i}\bra{X}_{t_i}}{\sqrt 2 } \to \frac{\bra{0}_{h_i}\bra{0}_{t_i}+ (- 1)^{f(X)}\bra{X}_{h_i}\bra{X}_{t_i}}{\sqrt 2 }\]
where $i=1,2$. They are then sent back to Alice.}
    \item {After receiving particles ${t_1}$ and ${t_2}$, Alice performs Shi et al.'s honesty test \cite{Shi2017}. First, she applied $2m$ times $CNOT$ gates on $2m$ pairs of entangled qubits, where ${t_i}$ controls ${h_i}$: 
\[\begin{aligned}
CNOT^{ \otimes m}:&\frac{\bra{0}_{h_i}+\bra{0}_{t_i}+ (- 1)^{f(X)}\bra{X}_{h_i}\bra{X}_{t_i}}{\sqrt{2}} \\
&\to \bra{0} _{h_i}\frac{\bra{0}_{t_i}+ (-1)^{f(X)}\bra{X}_{t_i}}{\sqrt{2} }.\end{aligned}\]
Then Alice measures ${h_i}$ in the calculation basis $\bra{j},$. If both ${h_i}$ measurements are $\bra{0}$, then the next step is carried out; otherwise, Bob is considered to be dishonest.}
\end{enumerate}

\subsubsection{Output stage}\label{sec3.2.3}

\begin{enumerate}[Step 1]
    \item {According to the measurement method in Section~\ref{sec2.2}, Alice performs projection measurement on the states in particle ${t_1}$ and ${t_2}$ respectively to distinguish the states 
\[\bra{X_+}= \frac{{\bra{0}_{t_i}}+ \bra{X}_{t_i}}{{\sqrt 2 }}\]
and 
\[\bra{X_-}= \frac{{\bra{0}_{t_i}}- \bra{X}_{t_i}}{{\sqrt 2 }}.\]
She then tests whether the measurement results of ${t_1}$ and ${t_2}$ are the same. If they are the same, goes to the next step; otherwise, considers Bob cheated.}

    \item {Alice determines the intersection of two circles according to whether her measurement result is 
$\bra{X_+}$ or $\bra{X_-}$. If it's $\bra{X_-}$, then the circles intersect, otherwise they don't.}

    \item Alice tells Bob the result through the classical channel.
\end{enumerate}

\section{Performance analysis}\label{sec4}
\subsection{Correctness}\label{sec4.1}
Obviously, when two circles intersect, we have
\[D = {\left( {{x_1} - {x_2}} \right)^2} + {\left( {{y_1} - {y_2}} \right)^2} < {\left( {{r_1} + {r_2}} \right)^2} = R.\]
Since the final measurement result is 
\[\frac{\bra{0}_{t_i}+ (- 1)^{D < R?}\bra{X}_{t_i}}{\sqrt 2 },\]
if $\bra{X_-}$ is measured, it means $D < R$, so the two circles intersect; Otherwise, if $\bra{X_+}$, then $D \ge R$, i.e., the circles don't intersect.

\subsection{Security}\label{sec4.2}

\noindent\textbf{Alice's privacy under Bob's attacks:}

\noindent(1) Direct measurement attack

If Bob directly measures particle ${t_i}$, state 
\[\bra{\psi_i} = \frac{\bra{0}_{h_i}\bra{0}_{t_i}+ \bra{X}_{h_i}\bra{X}_{t_i}}{\sqrt 2 }\]
will collapse into $\bra{X}_{h_i}\bra{X}_{t_i}$
with probability $\frac{1}{2}$, and then Bob obtains the information $X$, which is unavoidable. However, when Alice performs measurement on base $\bra{X_{\pm}}$ in step 2 of the output stage, due to 
\[\left\lvert X \right\rangle  = \frac{1}{{\sqrt 2 }}\frac{{\left\lvert 0 \right\rangle  + \left\lvert X \right\rangle }}{{\sqrt 2 }} - \frac{1}{{\sqrt 2 }}\frac{{\left\lvert 0 \right\rangle  - \left\lvert X \right\rangle }}{{\sqrt 2 }} = \frac{1}{{\sqrt 2 }}\left\lvert {{X_ + }} \right\rangle  - \frac{1}{{\sqrt 2 }}\left\lvert {{X_ - }} \right\rangle, \]
she will get a result randomly with probability $\frac{1}{2}$, and the probability that the two particles $t_1,t_2$ output the same results is $\frac{1}{2}$. That is, she will detect Bob's cheating behavior with half probability. Similarly, if the state collapses into ${\left\lvert 0 \right\rangle _{{h_i}}}{\left\lvert 0 \right\rangle _{{t_i}}}$, it will also be found with the same probability. In total, the probability of Bob to obtain effective information while concealing cheating behavior is $\frac{1}{4}$.

\noindent(2) Intercept-and-resend attack

If Bob not only performs measurement on particle ${t_i}$ after receiving it, but also does not send it back, but sends a fake particle ${e_i}$, then to conceal his cheating under the $CNOT^{\otimes m}$ operation between ${h_i}$ and ${t_i}$ in step 3 of the operation stage, he must ensure that the measurement values of the fake particle ${e_i}$ and ${t_i}$ are in the same state. Assume that the result state is  ${\left\lvert X \right\rangle _{{h_i}}}{\left\lvert X \right\rangle _{{t_i}}}{\left\lvert X \right\rangle _{{e_i}}}$, then after Alice apply $CNOT^{\otimes m}$, it will change to
\[{\left\lvert X \right\rangle _{{t_i}}}{\left\lvert X \right\rangle _{{e_i}}} = \left( {\frac{1}{{\sqrt 2 }}{{\left\lvert {{X_ + }} \right\rangle }_{{t_i}}} - \frac{1}{{\sqrt 2 }}{{\left\lvert {{X_ - }} \right\rangle }_{{t_i}}}} \right){\left\lvert X \right\rangle _{{e_i}}}.\]
Similarly, after Alice performs the measurement, it will collapse to ${\left\lvert {{X_ + }} \right\rangle _{{t_i}}}{\left\lvert X \right\rangle _{{e_i}}}$ or ${\left\lvert {{X_ - }} \right\rangle _{{t_i}}}{\left\lvert X \right\rangle _{{e_i}}}$ with half probability, and all conclusions are the same as those in the direct measurement attack.

\noindent(3) Entangle-and-measure attack

If Bob prepares an auxiliary particle ${e_i}$, and entangles ${t_i}$ and ${e_i}$ through the unitary operator ${\tilde U}:{\left\lvert j \right\rangle _{{t_i}}}{\left\lvert 0 \right\rangle _{{e_i}}} \to {\left\lvert j \right\rangle _{{t_i}}}{\left\lvert {\varepsilon \left( j \right)} \right\rangle _{{e_i}}}$: 
\[{\tilde U}:\frac{\bra{0}_{h_i}\bra{0}_{t_i} + \bra{X}_{h_i}\bra{X}_{t_i}}{\sqrt 2}\bra{0}_{e_i} \to 
\frac{\bra{0}_{h_i}\bra{0}_{t_i}\bra{\varepsilon(0)}_{e_i} + \bra{X}_{h_i}\bra{X}_{t_i}\bra{\varepsilon(X)}_{e_i}}{\sqrt 2}
\]
Then he keeps ${e_i}$ and sends ${t_i}$ back to Alice. He can pass the $CNOT^{\otimes m}$ test because ${t_i}$ hasn't changed. In this case, the total state is 
\[\begin{aligned}
&\frac{\bra{0}_{t_i}\bra{\varepsilon(0)}_{e_i} + \bra{X}_{t_i}\bra{\varepsilon(X)}_{e_i}}{\sqrt 2}\\
& = \frac{1}{{\sqrt 2 }}\left(\frac{\bra{X_+}_{t_i}+ \bra{X_-}_{t_i}}{\sqrt 2 } \bra{\varepsilon(0)}_{e_i} +\frac{\bra{X_+}_{t_i}- \bra{X_-}_{t_i}}{\sqrt 2 } \bra{\varepsilon(X)}_{e_i}\right)\\
& = \frac{1}{{\sqrt 2 }}\bra{X_+}_{t_i}\frac{\bra{\varepsilon(0)}_{e_i}+ \bra{\varepsilon(X)}_{e_i}}{\sqrt 2 }  + \frac{1}{{\sqrt 2 }}\bra{X_-}_{t_i}\frac{\bra{\varepsilon(0)}_{e_i}- \bra{\varepsilon(X)}_{e_i}}{\sqrt 2}.
\end{aligned}\]
After Alice's measurement, it collapses to 
\[\bra{X_+}_{t_i}\frac{\bra{\varepsilon(0)}_{e_i}+ \bra{\varepsilon(X)}_{e_i}}{\sqrt 2 }\]
with probability ${p_ + }$, or 
\[\bra{X_-}_{t_i}\frac{\bra{\varepsilon(0)}_{e_i}- \bra{\varepsilon(X)}_{e_i}}{\sqrt 2 }\]
with probability ${p_ - } = 1 - {p_ + }$. No matter what the result is, Bob can only get $\bra{\varepsilon(X)}_{e_i}$ with half probability, and thus determine the value of $X$. On the other hand, in order to fully distinguish between the two states $\bra{\varepsilon(0)}_{e_i}$ and $\bra{\varepsilon(X)}_{e_i}$, he must guarantee that these two are orthogonal, so that both $\frac{\bra{\varepsilon(0)}_{e_i}+ \bra{\varepsilon(X)}_{e_i}}{\sqrt 2 }$ and $\frac{\bra{\varepsilon(0)}_{e_i}- \bra{\varepsilon(X)}_{e_i}}{\sqrt 2 }$ will be unit vectors, and thus ${p_ + } = {p_ - } = \frac{1}{2}$. Then, all the analysis is consistent with the above.

To sum up, in all possible attacks, Bob can only get $X$ with $\frac{1}{2}$ probability, and Alice can detect the result with the same probability.

\noindent\textbf{Bob's privacy under Alice's attacks:}

\noindent(1) Multiple input attack

If Alice inputs different $X$ about the circle area in $h_1,t_1$ and $h_2,t_2$, she will be able to get two results at once. In fact, in the malicious attack model, this kind of attacks are generally impossible to defend against. Since the size of the grid is exponential, Alice can't perform exponential times this attack, and so she doesn't gain much advantage.

\noindent(2) Superimposed input attack

Suppose that the state Alice prepared is not the superposition state of $\left\lvert 0 \right\rangle$ and $\left\lvert X \right\rangle$, but the uniform superposition state $\frac{1}{{\sqrt N }}\sum_{j=0}^{N-1} {\left\lvert j \right\rangle }$. She wants to analyze Bob's private circle information at one time. Note that Bob's Oracle operator is the Oracle operator used in Grover's algorithm \cite{Grover1997}, an effective attack will be equivalent to this algorithm. However, Grover's algorithm needs to carry out $O\left( {\sqrt N } \right)$ consecutive times, so the attack is inefficient. On the other hand, she can only end up with one solution at most, because $m$-qubits can transmit $m$-bits of classical information at most \cite{Nielsen2010}. The solution Alice can get is exactly a circle that can intersect Bob's circle, but because there are different kinds of circles, the information she can get is very small. Therefore, such attacks are not effective.

\noindent\textbf{Alice and Bob's privacy under external attacks:}

Consider a external attacker Eve who has no knowledge about Alice and Bob's privacy. According to the analysis above, if Eve intercepts the particles sent by Alice to Bob, she is now similar to Bob exerting a direct measurement attack, and will be detected with probability $\frac{1}{2}$. If she intercepts the particles returned by Bob to Alice, it is impossible to get information about Bob, because she know nothing about $X$ so that she can't distinguish $\bra{X_+}$ and $\bra{X_-}$. In addition, to prevent such an eavesdroppers, Alice and Bob can also perform eavesdropper testing, i.e., inserting a number of decoy particles in a specified state into the particles $t_i$ and verifying whether they have been measured.

\subsection{Complexity}\label{sec4.3}
First we consider the complexity of the Oracle operator $U$ described in Section~\ref{sec3.1}. Since the computational complexity of all  operations is at most $O\left( {{n^2}} \right)$, as mentioned in Section~\ref{sec2.3}, the total computational complexity is also this much. Note that $n = 2t + 3=O(t)$, so the complexity is $O\left( {{t^2}} \right)$.

As described in Section~\ref{sec2.2}, the preparation of the superposition (the preparation stage) and the distinguishing of results (the output stage) both have computational complexity of $O\left( {{m}} \right)=O\left( {{t}} \right)$. In total, our protocol has computational complexity of $O\left( {{t^2}} \right)$ and qubit complexity of $O\left( t \right)$. Since in the protocol we need transmit several $m = O\left( t \right)$ qubits of quantum information, the communication complexity is also $O\left( {{t}} \right)$. Therefore, our protocol has polynomial complexity and is efficient.

\section{Conclusion}\label{sec5}

In this paper, we use the principle of phase-encoded query to realize the quantum privacy-preserving circle intersection decision. We decompose the Oracle operator used in the query into quantum arithmetic operations in detail, so as to achieve polynomial computational complexity, and avoid the problem that the Oracle operator is difficult to implement in the existing protocols based on phase-encoded query. Performance analysis shows that our protocol is correct and efficient, and can protect the privacy of all participants against internal and external attacks. Our protocol gives a new way for the development of Quantum privacy-preserving computational geometry.

\bmhead{Acknowledgments}

This work is supported by the National Natural Science Foundation of China (62071240), the Innovation Program for Quantum Science and Technology (2021ZD0302900), and the Priority Academic Program Development of Jiangsu Higher Education Institutions (PAPD).

\section*{Declarations}

\begin{itemize}
\item \textbf{Conflict of interest} The authors declare that they have no conflict of interest.
\item \textbf{Ethical statement} Articles do not rely on clinical trials.
\item \textbf{Human and animal participants} All submitted manuscripts containing research which does not involve human participants and/or animal experimentation.

\item \textbf{Data availability} Data sharing not applicable to this article as no datasets were generated or analysed during the current study.
\end{itemize}


\begin{thebibliography}{50}
\bibitem{Yao1982}Yao,A.C.:Protocols for secure computations.In:Proceeding of 23rd IEEE Symposium on Foundations of Computer Science,pp.160-164.IEEE,Piscataway(1982).https://doi.org/10.1109/SFCS.1982.38

\bibitem{Shi2015} Shi, R.H., Mu, Y., Zhong, H., Zhang, S.: Quantum oblivious set-member
decision protocol. Physical Review A \textbf{92}(2), 022309 (2015). https://doi.
org/10.1103/PhysRevA.92.022309

\bibitem{Ji2019} Ji, Z.X., Zhang, H.G., Wang, H.Z., Wu, F.S., Jia, J.W., Wu,
W.Q.: Quantum protocols for secure multi-party summation. Quan-
tum Information Processing \textbf{18}(6), 168 (2019). https://doi.org/10.1007/
s11128-018-2141-1

\bibitem{Liu2019B} Liu, W.J., Li, C.T., Zheng, Y., Xu, Y., Xu, Y.S.: Quantum
privacy-preserving price e-negotiation. International Journal of The-
oretical Physics \textbf{58}(10), 3259–3270 (2019). https://doi.org/10.1007/
s10773-019-04201-9

\bibitem{Shi2021} Shi, R.H., Liu, B., W., Z.M.: Secure two-party integer comparison protocol
without any third party. Quantum Information Processing \textbf{20}(12), 402
(2021). https://doi.org/10.1007/s11128-021-03344-1

\bibitem{Liu2022} Liu, W.J., Li, W.B., Wang, H.B.: An improved quantum private set intersection protocol based on hadamard gates. International Journal of Theoretical Physics \textbf{61}(3), 53 (2022). https://doi.org/10.1007/s10773-022-05048-3

\bibitem{Ye2022} Ye, T.Y., Xu, T.J., Geng, M.J., Chen, Y.: Two-party secure semi-quantum summation against the collective-dephasing noise. Quantum Information Processing \textbf{21}(3), 118 (2022). https://doi.org/10.1007/s11128-022-03459-z

\bibitem{Atallah2001}
Atallah,M.J.,Du,W.:Secure Multi-party Computational Geometry.In:Dehne,F,J.Sack,J.-R.,Tamassia,R.(eds)Algorithms and Data Structures,pp.165--179.Springer Berlin Heidelberg(2001).https://doi.org/10.1007/3-540-44634-6-16

\bibitem{Huang2016}Huang, H., Gong, T., Chen, P., Malekian, R., Chen, T.: Secure two-party distance computation protocol based on privacy homomorphism
and scalar product in wireless sensor networks. Tsinghua Science and
Technology \textbf{21}(4), 385–396 (2016). https://doi.org/10.1109/TST.2016.7536716

\bibitem{Chen2018}Chen, B.R., Yang, W., Huang, L.S.: Cryptanalysis and improvement of the novel quantum scheme for secure two-party distance computation.
Quantum Information Processing \textbf{18}(1),35(2018). https://doi.org/10.1007/s11128-018-2148-7

\bibitem{Liu2019}Liu, W.J., Xu, Y., Yang, J.C.N., Yu, W.B., Chi, L.H.: Privacy-preserving quantum two-party geometric intersection. Computers, Materials \& Con-tinua \textbf{60}(3), 1237–1250 (2019). https://doi.org/10.32604/cmc.2019.03551

\bibitem{Shi2017}Shi,R.H.,Mu, Y., Zhong,H.,Cui,J.,Zhang,S.:Privacy-preserving point-
inclusion protocol for an arbitrary area based on phase-encoded quantum
private query. Quantum Information Processing \textbf{16}(1),8(2017). https://doi.org/10.1007/s11128-016-1476-8

\bibitem{Peng2017}Peng,Z.W.,Shi, R.H.,Zhong,H.,Cui,J.,Zhang,S.:A novel quantum scheme for secure two-party distance computation.Quantum
Information Processing \textbf{16}(12),316 (2017).https://doi.org/10.1007/s11128-017-1766-9

\bibitem{Peng2018} Peng, Z.W., Shi, R.H., Wang, P.H., Zhang, S.: A novel quantum solution
to secure two-party distance computation. Quantum Information Process-ing \textbf{17}(6),145 (2018). https://doi.org/10.1007/s11128-018-1911-0


\bibitem{Cao2022} Cao, Y.H.: Quantum secure two-party euclidean distance computation
based on mutually unbiased bases. Quantum Information Processing
\textbf{21}(7), 262 (2022). https://doi.org/10.1007 /s11128-022-03611-9

\bibitem{He2012} He, L.B., Huang, L.S., Yang, W., Xu, R.: A protocol for the secure two-
party quantum scalar product. Physics Letters A \textbf{376}(16), 1323–1327
(2012). https://doi.org/10.1016/j.physleta.2012.02.048

\bibitem{Shi2019}Shi, R.H., Zhang, M.W.: Strong privacy-preserving two-party scalar
product quantum protocol. International Journal of Theoretical Physics
\textbf{58}(12), 4249–4257 (2019). https://doi.org/10.1007/s10773-019-04296-0

\bibitem{Li2014}Li, S.D., Wu, C.Y., Wang, D.S., Dai, Y.Q.: Secure multiparty computation
of solid geometric problems and their applications. Information Sciences
\textbf{282}, 401–413 (2014). 2014). https://doi.org/10.1016/ j.ins.2014.04.004


\bibitem{Zhu2018} Zhu, H., Wang, F., Lu, R., Liu, F., Fu, G., Li, H.: Efficient and privacy-
preserving proximity detection schemes for social applications. IEEE
Internet of Things Journal \textbf{5}(4), 2947–2957 (2018). https://doi.org/10.
1109/JIOT.2017.2766701

\bibitem{Olejnik2011} Olejnik, L.: Secure quantum private information retrieval using phase-
encoded queries. Physical Review A \textbf{84}(2), 022313 (2011). https://doi.
org/10.1103/PhysRevA.84.022313

\bibitem{Grover1997} Grover, L.K.: Quantum mechanics helps in serching for a needle in a
haystack. Physical Review Letters \textbf{79}(2), 325–328 (1997). https://doi.org/
10.1103/PhysRevLett.79.325

\bibitem{Shor1997} Shor, P.W.: Polynomial-time algorithms for prime factorization and dis-
crete logarithms on a quantum computer. Siam Journal on Computing
\textbf{26}(5), 1484–1509 (1997). https://doi.org/10.1137/S0097539795293172

\bibitem{Nielsen2010} Nielsen, M.A., Chuang, I.L.: Quantum Computation and Quantum Infor-
mation: 10th Anniversary Edition. Cambridge University Press, New York
(2010). https://doi.org/10.1017/CBO9780511976667
\end{thebibliography}
\end{document}